\documentclass[twocolumn,amsmath,preprintnumbers,superscriptaddress,amssymb]{revtex4-1}
\usepackage{graphicx}
\usepackage{xcolor}
\usepackage[T2A]{fontenc}

\begin{document}
\title{Dissipation entanglement control of two coupled qubits via strong driving fields}

\author{M. V. Bastrakova}
\email{bastrakova@phys.unn.ru}
\author{V. O. Munyaev}

\affiliation{Lobachevsky State University of Nizhny Novgorod, 23 Gagarin Ave., 603022, Nizhny Novgorod, Russia}

\begin{abstract}
Abstract.
An analytical theory to calculate the dissipatively stable concurrence in the system of two coupled flux superconducting qubits in the strong driving field is developed. The conditions for the entanglement state generation and destruction during the formation of the multiphoton transitions regions due to the interference of Landau--Zener--St\"uckelberg--Majorana are found. Based on the solution of the Floquet--Markov equation, the technique is proposed to adjust the amplitudes of dc- and ac-fields for effective control of the entanglement between qubit states while taking into account the effects of the decoherence.
\end{abstract}

\date{\today}
\maketitle

\section{Introduction}
Quantum computing, quantum error correction procedures and quantum cryptography protocols are based on the principles of generation and high-precision control of entangled states of quantum objects~\cite{Horodecki2009,Zou2021}. During the last few years, one of the actively developed concepts to manipulate quantum information is based on the use of environmental noise~\cite{Tacchino2018,Harrington2022}. The possibility of dissipatively entangled long-lived steady-states formation,  their transport and stabilization have been  already demonstrated in many physical systems, such as cavity quantum electrodynamics~\cite{Reiter2012,Blais2021,Gallardo2022}, atomic ensembles \cite{Krauter2011, Santos2022}, trapped ions~\cite{Bentley2014, Cole2022}, quantum dots~\cite{Hichri2004,Anton2022}, and superconducting circuits~\cite{Kimchi-Schwartz2016,Campagne-Ibarcq2018}.

Entangled states creation in combination with relaxation rates control of quantum systems by weak resonance effects is successfully used today in practice in many important quantum-information applications~\cite{Kraus2008,Reiter2013,Kimchi-Schwartz2016,Cacciapuoti2020}: quantum algorithms' realization, quantum neural networks design, effective quantum communication protocol development, quantum transport  studying, creation of long-lived qubit-resonator states, etc.
In addition, a protocol for generating stationary dissipative entanglement in two coupled superconductor qubits based on non-resonant periodic driving of large amplitude has been recently proposed~\cite{Gramajo2018,Bonifacio2020,Gramajo2021}. This approach  allows studying of interesting non-perturbative quantum effects in superconductor qubits, such as Landau--Zener--St\"uckelberg--Majorana interference (LZSM) \cite{Berns2006, Izmalkov2008, Shevchenko2010, Neilinger2016, Munyaev2021, Ivakhnenko2023}. 
In addition, LZSM interferometry was recently proposed to determine relevant information related to the coupling of the qubit with a noisy environment~\cite{Shevchenko2010,Forster2014,Blattmann2015,Mi2018,Gramajo2019,Munyaev2021,Ivakhnenko2023}. 

Initially, studies of LZSM dynamics and entanglement formation processes in two coupled superconducting qubits were carried out in the framework of the quasi-energetic representation and the Floquet formalism~\cite{Shirley1965,Grifoni1998,Zeldovich1967,Ritus1967,Sambe1973} under the condition of high coherence, i.e., when the decoherence time of the qubits is much larger than the time of periodic external influence on the system. Using these approaches in the case of a strong driving field, the conditions for the formation of multiphoton resonances were found~\cite{Ilichev2010,Satanin2012,Munyaev2021}, the induction of stationary entanglement in two qubits was shown and the process of its control by external fields and the coupling force between the qubits was demonstrated~\cite{Zhang2009,Gramajo2018}. Further, the formation of dissipatively stable steady-state in the system of two coupled superconductor qubits at times greater than the decoherence time under the condition of a strong alternating field was numerically detected in the works~\cite{Bonifacio2020,Gramajo2021,Munyaev2021}. However, the questions about the analytical description of the correspondence between the regions of multiphoton resonances and the realization of control by the external field of long-lived entangled qubit states remain unclear.

In our early works~\cite{Satanin2012,Munyaev2021}, we developed an analytical description of the features of multiphoton transition regions formation without taking into account the influence of dissipation in the two coupled flux qubits system. In addition, in the paper~\cite{Munyaev2021} we provide numerical estimates of the dissipative concurrence behavior and find its interesting effects near resonances that have not been previously described in the literature. In this connection, the present work is a logical continuation of the studies started in~\cite{Munyaev2021} and is devoted to a detailed theoretical analysis of the conditions for the formation and destruction of entanglement in the system of two coupled qubits during the formation of multiphoton transition regions as a result of LZSM interference. Here we investigate the steady-state entanglement and coherent dissipative dynamics  of two  inductively coupled qubits, each interacting with a local boson reservoir  using the basic Floquet--Markov equation~\cite{Kohler1997,Grifoni1998,Hausinger2010}.
The physical system model under study is described in Sec.~II. We focus on the strong driving field regime and study the concurrence behavior within the framework of Floquet perturbation theory for the master equation up to second order, based on small parameters of qubits tunnel splittings. First, in Sec.~III we carry out a deep analytical investigation of the non-resonant case and obtain an explicit expression for the averaged concurrence of two coupled qubits under the conditions of interaction with the bosonic reservoirs. Second, in Sec.~IV and V we develop the resonance perturbation theory and find the corresponding expressions for concurrence magnitude. In addition, we determined the conditions of entanglement generation in the rotating wave approximation, made estimates of the shape and width of the resonance curves based on resonance perturbation theory as a function of the dephasing and dissipation rates of the system. Section~VI and the conclusion present the results and show a good agreement between the developed theory and direct numerical simulations, and discuss the perspectives and possibilities for its further use and development.



\section{System model under study}
We get the Floquet--Markov master equation~\cite{Kohler1997,Grifoni1998,Hausinger2010} for two coupled flux qubits:
\begin{equation}
\frac{\partial\hat{\rho}}{\partial t}=-i\left[\hat{H}\left(t\right),\hat{\rho}\right]+\hat{\Gamma}\hat{\rho},
\label{eq:master_equation}
\end{equation}
with the global Hamiltonian:
\begin{equation}
\begin{gathered}
\hat{H}\left(t\right)=-\frac{1}{2}\sum_{q=1}^{2}\left(\epsilon_q(t)\sigma_z^{\left(q\right)}+\Delta_q\sigma_x^{\left(q\right)}\right)-\frac{g}{2}\sigma_z^{\left(1\right)}\sigma_z^{\left(2\right)},
\end{gathered}
\label{eq:hamiltonian_global}
\end{equation}
where $\sigma_z^{\left(q\right)}$, $\sigma_x^{\left(q\right)}$ are the Pauli matrices, with $q=1,2$ the index of each qubit. We take $ \hbar= 1$ in this work. The parameter $\Delta_q$ is the tunnel splitting of the qubit levels: \textcolor{black}{$\left|\downarrow^{\left(q\right)}\right\rangle$ (ground state) and  $\left|\uparrow^{\left(q\right)}\right\rangle$ (excited state), which defines the computational basis. The strength of the interaction between qubits is defined as $g$. The qubit is driven by a time dependent bias $\epsilon_q(t) = \epsilon_q +v_{q}\left(t\right)$, where $\epsilon_q$ is the static bias component of external field, and $v_{q}\left(t\right) = A_{q}\cos\left(\omega t-\varphi_0\right)$ is the harmonic variable part of the magnetic flux of the microwave amplitude field, $A_{q}$, and frequency, $\omega$, applied to each qubit. We assume that the system is under the influence of a pulse sequence of the alternating fields with fixed period $T=2\pi/\omega$. At the same time, there may occur losses and phase shifts of the pulses when passing through the coaxial lines, which will affect the arrival time of the pulse on the qubits. To account for this effect, let’s denote the random time, $t_0$, of the pulse arrival on the qubits or the corresponding random phase $\varphi_0=\omega t_0$. }


The dissipative operator in Eq.~\eqref{eq:master_equation} is defined by:
\begin{equation}
\hat{\Gamma}=\sum_{q=1}^{2}\!\left(\Gamma_{\varphi_q}\hat{D}\!\left[\hat{\sigma}_{z}^{\left(q\right)}\right]+\Gamma_q\hat{D}\!\left[\hat{\sigma}_{-}^{\left(q\right)}\right]+\Gamma^\prime_{q}\hat{D}\!\left[\hat{\sigma}_{+}^{\left(q\right)}\right]\right)\!,
\label{eq:operator_dissipative}
\end{equation}
where $\Gamma_{\varphi_q}$, $\Gamma_q$ and $\Gamma^\prime_q$ are dephasing, relaxation and excitation incoherent rates, respectively; $\hat{D}\!\left[\hat{a}\right]\hat{\rho}\equiv\hat{a}\hat{\rho}\hat{a}^{\dag}-\frac{1}{2}\left\{\hat{a}^{\dag}\hat{a},\hat{\rho}\right\}$ and the Lindblad operators $\hat{\sigma}_z^{\left(q\right)}$, $\hat{\sigma}_{+}^{\left(q\right)}$, $\hat{\sigma}_{-}^{\left(q\right)}$ are expressed via computational basis (as if there were no coupling, $g=0$, and field, $v_{q}\left(t\right) = 0$, in Eq.~\eqref{eq:hamiltonian_global}) as follows
\begin{equation}
\begin{aligned}
&\hat{\sigma}_z^{\left(q\right)}=\left|\uparrow^{\left(q\right)}\right\rangle\left\langle\uparrow^{\left(q\right)}\right|-\left|\downarrow^{\left(q\right)}\right\rangle\left\langle\downarrow^{\left(q\right)}\right|,\\
&\hat{\sigma}_{+}^{\left(q\right)}=\left|\uparrow^{\left(q\right)}\right\rangle\left\langle\downarrow^{\left(q\right)}\right|,
\;\hat{\sigma}_{-}^{\left(q\right)}=\left|\downarrow^{\left(q\right)}\right\rangle\left\langle\uparrow^{\left(q\right)}\right|.
\end{aligned}
\end{equation}
Note that the Eq.~\eqref{eq:master_equation} is valid in the case the incoherent rates are much smaller than all coherent energy scales (approximation of independent rates~\cite{Cohen1998}). At the reservoirs fundamental temperature $\tau_B$, the relaxation and excitation parameters are related as $\Gamma^\prime_q=\Gamma_q\exp\!\left(-\Delta E^{\left(q\right)}/\tau_B\right)$, where $\Delta E^{\left(q\right)}$ is the energy gap of the $q$-th qubit. At higher temperatures, it is necessary to account for the effect of the Hamiltonian’s time dependence on the matrix elements of the qubit transitions~\cite{Xu2014}.

\section{Concurrence in the nonresonant case}
In our previous work~\cite{Munyaev2021}, the multiphoton transitions in a non-dissipative system of two coupled qubits with Hamiltonian~\eqref{eq:hamiltonian_global} have been studied in detail for the case of small qubit tunneling splitting $\Delta_q\ll{A},\epsilon_q$ and $A_1=A_2=A$, $\varphi_0=0$. According to the Floquet theorem~\cite{Shirley1965, Grifoni1998}, due to the Hamiltonian periodicity, $\hat{H}\!\left(t\right)=\hat{H}\!\left(t+T\right)$, the solution of the corresponding Schr\"{o}dinger equation can be spanned in the Floquet basis~\cite{Zeldovich1967,Ritus1967,Sambe1973}. The general solution of the Schr\"{o}dinger equation for an arbitrary phase $\varphi_0$ can be expressed through $\left|u_{\alpha}\left(t\right)\right>$ and  $\gamma_{\alpha}$ as follows:
\begin{equation}
  \left|\varPsi\left(t\right)\right>=\sum_{\alpha}c_{\alpha}e^{-i\gamma_{\alpha}t}\left|u_{\alpha}\left(t-\varphi_0/\omega\right)\right>,
\label{eq:psi_solution_general}
\end{equation}
where coefficients $c_{\alpha}$ are defined from the initial condition. The periodic Floquet modes $\left\{\left|u_\alpha\left(t\right)\right>\right\}$, $\left|u_{\alpha}\left(t+T\right)\right>=\left|u_{\alpha}\left(t\right)\right>$, and corresponding quasienergies $\gamma_{\alpha}$ ($\alpha = 1,2,3,4$ is the quantum number determining the quasienergy for two coupled qubits) of the system~\eqref{eq:hamiltonian_global} were found perturbatively for the case of small qubit tunneling splitting $\Delta_q\ll{A},\epsilon_q$ and $A_1=A_2=A$, $\varphi_0=0$ (see~\cite{Munyaev2021}).

Floquet states and quasienergies $\pmod\omega$ in zeroth order, $\Delta_1=\Delta_2=0$,
are, respectively,
\begin{equation}
\begin{aligned}
&\left|u_{1}^{\left(0\right)}\!\left(t\right)\right>\!=\!\Big(\!e^{i\frac{A}{\omega}\sin\omega t},0,0,0\!\Big)^T\!,&&\gamma_{1}^{\left(0\right)}\!=\!-\frac{\epsilon_{1}\!+\!\epsilon_{2}\!+\!g}{2},\\
&\left|u_{2}^{\left(0\right)}\!\left(t\right)\right>\!=\!\Big(\!0,1,0,0\!\Big)^T\!,&&\gamma_{2}^{\left(0\right)}\!=\!-\frac{\epsilon_{1}\!-\!\epsilon_{2}\!-\!g}{2},\\
&\left|u_{3}^{\left(0\right)}\!\left(t\right)\right>\!=\!\Big(\!0,0,1,0\!\Big)^T\!,&&\gamma_{3}^{\left(0\right)}\!=\!\frac{\epsilon_{1}\!-\!\epsilon_{2}\!+\!g}{2},\\
&\left|u_{4}^{\left(0\right)}\!\left(t\right)\right>\!=\!\Big(\!0,0,0,e^{-i\frac{A}{\omega}\sin\omega t}\!\Big)^T\!,&&\gamma_{4}^{\left(0\right)}\!=\!\frac{\epsilon_{1}\!+\!\epsilon_{2}\!-\!g}{2}.
\end{aligned} \label{eq:floquet_solution_unperturbed}
\end{equation}

The analytical results obtained in~\cite{Munyaev2021} can be used for preliminary analysis of the behavior and computation of the system~\eqref{eq:master_equation} concurrence in the limit of small incoherent rates $\Gamma_{\varphi_q}$, $\Gamma_q$ and $\Gamma'_q$. Indeed, for sufficiently small incoherent rates the factorizing approximation $\hat{\rho}\approx\left|\psi\right>\left<\psi\right|$ can be used, where the state $\left|\psi\right>$ are defined by Eq.~\eqref{eq:psi_solution_general}. To take into account dissipative effects on the Hamiltonian solution~\eqref{eq:psi_solution_general} we consider the simple case of zero fundamental temperature $\tau_B=0$ ($\Gamma'_1=\Gamma'_2=0$). It is easy to show that any solution $\hat{\rho}$ of~\eqref{eq:master_equation} in the lowest perturbation theory order, that is, for $\Delta_1=\Delta_2=0$, tends to $\hat{\rho}\to\left|\downarrow\downarrow\right>\!\left<\downarrow\downarrow\right|$. Comparing this result with~\eqref{eq:psi_solution_general} and~\eqref{eq:floquet_solution_unperturbed} we conclude that $c_2=c_3=c_4=0$ in Eq.~\eqref{eq:psi_solution_general}. Thus, at small incoherent rates we obtain:
\begin{equation}
  \hat{\rho}\left(t\right)\approx\left|u_{1}\left(t-\varphi_0/\omega\right)\right>\!\left<u_{1}\left(t-\varphi_0/\omega\right)\right|.
\label{eq:rho_solution_approx}
\end{equation}

We calculate the average entanglement measure as a concurrence~\cite{Wootters1998}: $C\left(\rho\right)=\max\left\{0,\lambda_4-\lambda_3-\lambda_2-\lambda_1\right\}$, where $\lambda_i$'s are real numbers in decreasing order and correspond to the eigenvalues of the matrix $R=\sqrt{\sqrt{\rho}\Tilde{\rho}\sqrt{\rho}}$, with $\Tilde{\rho}=\sigma^{\left(1\right)}_y\otimes\sigma^{\left(2\right)}_y\rho^{*}\sigma^{\left(1\right)}_y\otimes\sigma^{\left(2\right)}_y$. Taking into account $\left|u_1\left(t\right)\right>=\sum\limits_{k}e^{ik\omega{t}}\left|u_{1k}\right>$, where
\begin{widetext}
\begin{equation}
\begin{gathered}
\left|u_{1k}\right>\!=\!
\begin{pmatrix} J_k\left(\frac{A}{\omega}\right)\left(1\!-\!\dfrac{1}{2}\sum\limits_{n=-\infty}^{+\infty}\left(\Delta_1^2\lambda_{1 n}^{2}\!+\!\Delta_2^2\lambda_{2 n}^{2}\right)\right)\!+\!\dfrac{1}{2}\sum\limits_{\substack{m=-\infty\\m\ne0}}^{+\infty}J_{k-m}\left(\frac{A}{\omega}\right)\dfrac{\Delta_1^2\chi_{1,-m}\!+\!\Delta_2^2\chi_{2,-m}}{m\omega}\\
\Delta_2\lambda_{2 k}\\
\Delta_1\lambda_{1 k}\\
\dfrac{\Delta_1\Delta_2}{2}\sum\limits_{n,m=-\infty}^{+\infty}\left(\lambda_{1 n}\!+\!\lambda_{2 n}\right)\dfrac{J_{m-k}\left(\frac{A}{\omega}\right)J_{m-n}\left(\frac{A}{\omega}\right)}{\epsilon_1\!+\!\epsilon_2\!+\!m\omega}
\end{pmatrix}, \\
\end{gathered}
\end{equation}
\end{widetext}
with parameters $\lambda_{qk}=\frac{J_k\left(A/\omega\right)}{2\left(\epsilon_{q}+g+k\omega\right)}$ and $\chi_{qk}=\sum\limits_{n=-\infty}^{+\infty}J_{n+k}\left(\frac{A}{\omega}\right)\lambda_{qn}$. Using Eq.~\eqref{eq:rho_solution_approx} we find the explicit form of the main term of the
infinite series in the smallness parameters $\Delta_q$ for concurrence $C$:
\begin{gather}
C\left(t\right)=2\Delta_{1}\Delta_{2}\left|{\sum\limits_{k=-\infty}^{+\infty}\!C_{k}e^{ik\left(\omega{t}-\varphi_0\right)}}\right|\!, \label{eq:C_result_no_resonance}\\
C_k\!=\!\sum\limits_{n=-\infty}^{+\infty}\!\left[\left(\lambda_{1n}+\lambda _{2n}\right)\!\frac{J_{k-n}\left(A/\omega\right)}{2\left(\epsilon_1+\epsilon_2+k\omega\right)}\!-\!\lambda_{1n}\lambda_{2,k-n}\right]\!. \notag
\end{gather}
The expression~\eqref{eq:C_result_no_resonance} shows that the resonance behavior of concurrence in the found approximation is observed when one of the following three conditions is met
\begin{subequations}
\begin{align}
\epsilon_{1,2}+g+k\omega\approx0, \label{eq:resonance_condition_a}\\
\epsilon_1+\epsilon_2+k\omega\approx0, \label{eq:resonance_condition_b}
\end{align}
\end{subequations}
which exactly coincide with the resonance conditions of transitions $1\to2$, $1\to3$ and $1\to4$ found earlier in~\cite{Munyaev2021}. Averaging of Eq.~\eqref{eq:C_result_no_resonance} over the pulse length $\tau$ and initial phase $\varphi_0$ leads to
\begin{equation}
\overline{C}=\frac{\Delta_{1}\Delta_{2}}{\pi}\int\limits_{0}^{2\pi}\!\mathrm{d}\alpha\!\left|{\sum\limits_{k=-\infty}^{+\infty}\!C_{k}e^{i{k}\alpha}}\right|\!.
\label{eq:C_result_no_resonance_average}
\end{equation}

\section{Rotating wave approximation}
We will assume that $\Delta_q\sim\lambda$ and $\Gamma_{\varphi_q},\Gamma_q,\Gamma^\prime_q\sim\lambda^2$, where $\lambda$ is the formal smallness parameter. Then, considering the Eq.~\eqref{eq:master_equation} with precision to terms of order~$\lambda^2$, the Lindblad operators can be chosen to be of order zero. When $\epsilon_{1,2}>0$:
\begin{equation}
\begin{aligned}
\hat{\sigma}_z^{\left(q\right)}\approx-\sigma_z^{\left(q\right)},\quad
\hat{\sigma}_{+}^{\left(q\right)}\approx\sigma_{-}^{\left(q\right)},\quad
\hat{\sigma}_{-}^{\left(q\right)}\approx\sigma_{+}^{\left(q\right)},
\end{aligned}
\end{equation}
where $\sigma_{\pm}^{\left(q\right)}\equiv\left(\sigma_x^{\left(q\right)}\pm i\sigma_y^{\left(q\right)}\right)/2$, known as ladder operators. The global Hamiltonian Eq.~\eqref{eq:hamiltonian_global} can be divided into an unperturbed part $\hat{H}_0\left(t\right)$ and a perturbing term $\hat{H}_1$
\begin{equation}
\begin{gathered}
\hat{H}\left(t\right)=\hat{H}_0\left(t\right)+\hat{H}_1,
\end{gathered}
\label{eq:hamiltonian_perturbation_division}
\end{equation}
where
\begin{gather}
\hat{H}_0\left(t\right)=-\frac{1}{2}\sum_{q=1}^{2}\epsilon_q(t)\sigma_z^{\left(q\right)}-\frac{g}{2}\sigma_z^{\left(1\right)}\sigma_z^{\left(2\right)},
\label{eq:hamiltonian_unperturbed_part}\\
\hat{H}_1=-\frac{1}{2}\sum_{q=1}^{2}\Delta_q\sigma_x^{\left(q\right)}.
\label{eq:hamiltonian_perturbed_part}
\end{gather}
First, we switch to the interaction representation, $\hat{\rho}=\hat{U}_0\hat{\rho}_I\hat{U}_0^{\dag}$, with respect to the unperturbed Hamiltonian~\eqref{eq:hamiltonian_unperturbed_part} for which the corresponding evolution operator $\hat{U}_0\left(t\right)$ is as follows:
\begin{align}
\hat{U}&_0\left(t\right)\!=\!\exp\left\{-i\:\mathrm{diag}\left[\gamma_1^{\left(0\right)}t\!-\!\frac{A}{\omega}\left(\sin\left(\omega t\!-\!\varphi_0\right)\!+\!\sin\varphi_0\right),\right.\right.\nonumber\\ 
&\left.\left.\gamma_2^{\left(0\right)}t,\gamma_3^{\left(0\right)}t,\gamma_4^{\left(0\right)}t\!+\!\frac{A}{\omega}\left(\sin\left(\omega t\!-\!\varphi_0\right)\!+\!\sin\varphi_0\right)\right]\right\}.
\label{eq:evolution_operator_unperturbed}
\end{align}
The interaction Hamiltonian $\hat{H}_{1,I}\left(t\right)=\hat{U}_0^{\dag}\hat{H}_1\hat{U}_0$ can then be shown to be
\begin{equation}
\begin{gathered}
\hat{H}_{1,I}\left(t\right)=-\frac{1}{2}\begin{pmatrix}
0 & \Delta_2\xi_2^{+*} & \Delta_1\xi_1^{+*} & 0\\
\Delta_2\xi_2^{+} & 0 & 0 & \Delta_1\xi_1^{-*}\\
\Delta_1\xi_1^{+} & 0 & 0 & \Delta_2\xi_2^{-*}\\
0 & \Delta_1\xi_1^{-} & \Delta_2\xi_2^{-} & 0
\end{pmatrix},
\end{gathered}
\label{eq:hamiltonian_interaction_representation}
\end{equation}
where the functions $\xi_q^{\pm}\left(t\right)$ are defined by
\begin{equation}
\begin{gathered}
\xi_q^{\pm}\!=\!\exp\!\left[i\!\left(\epsilon_q\pm g\right)\!t+i\frac{A}{\omega}\!\left(\sin\left(\omega t-\varphi_0\right)+\sin\varphi_0\right)\right]\!.
\end{gathered}
\label{eq:xi_q_pm_t}
\end{equation}

The dissipative operator Eq.~\eqref{eq:operator_dissipative} in the interaction representation looks like:
\begin{equation}
\hat{\Gamma}_I\!=\!\sum_{q=1}^{2}\!\left(\Gamma_{\varphi_q}\hat{D}\!\left[\hat{\sigma}_{z,I}^{\left(q\right)}\right]\!+\!\Gamma_q\hat{D}\!\left[\hat{\sigma}_{-,I}^{\left(q\right)}\right]\!+\!\Gamma^\prime_{q}\hat{D}\!\left[\hat{\sigma}_{+,I}^{\left(q\right)}\right]\right)\!,
\label{eq:operator_dissipative_interaction}
\end{equation}
where the Lindblad operators up to the zero order of the small parameters $\Delta_{1,2}$ and the insignificant phase factors are
\begin{align}
\hat{\sigma}_{z,I}^{\left(q\right)}=&\sigma_{z}^{\left(q\right)},\quad \hat{\sigma}_{+,I}^{\left(1\right)}=\sigma_{-}^{\left(1\right)}e^{i g\sigma_z^{\left(2\right)}t}, \quad \hat{\sigma}_{+,I}^{\left(2\right)}=e^{i g\sigma_z^{\left(1\right)}t}\sigma_{-}^{\left(2\right)},  \notag\\
&\hat{\sigma}_{-,I}^{\left(1\right)}=\sigma_{+}^{\left(1\right)}e^{-i g\sigma_z^{\left(2\right)}t},\quad
\hat{\sigma}_{-,I}^{\left(2\right)}=e^{-i g\sigma_z^{\left(1\right)}t}\sigma_{+}^{\left(2\right)}. \notag
\end{align}
So, Eq.~\eqref{eq:master_equation} with precision $\sim\lambda^2$ is transformed to the form
\begin{equation}
\frac{\partial\hat{\rho}_I}{\partial t}=-i\left[\hat{H}_{1,I}\left(t\right),\hat{\rho}_I\right]+\hat{\Gamma}_I\left(t\right)\hat{\rho}_I.
\label{eq:master_equation_interaction_representation}
\end{equation}
To be able to go further, we apply additional unitary transformation to Eq.~$\eqref{eq:master_equation_interaction_representation}$ with a unitary operator $\hat{U}_{1,I}\left(t\right)=\exp\left(-i\int_{0}^{t}\text{d}\tau\hat{H}_{1,I}\left(\tau\right)\right)$: $\hat{\rho}_I=\hat{U}_{1,I}\hat{\rho}_{II}\hat{U}_{1,I}^{\dag}$. Under this change, up to the second order in smallness parameters $\Delta_{1,2}$ the Hamiltonian $\hat{H}_{1,I}\left(t\right)$ transforms into $\hat{H}_{2,I}\left(t\right)=\frac{i}{2}\int_{0}^{t}\text{d}\tau\left[\hat{H}_{1,I}\left(\tau\right),\hat{H}_{1,I}\left(t\right)\right]$. Since $\hat{\Gamma}_I\left(t\right)$ already has the second order of smallness, for the chosen accuracy it does not change under the same transformation. So,
\begin{equation}
\frac{\partial\hat{\rho}_{II}}{\partial t}=-i\left[\hat{H}_{2,I}\left(t\right),\hat{\rho}_{II}\right]+\hat{\Gamma}_I\left(t\right)\hat{\rho}_{II}.
\label{eq:master_equation_interaction_representation_2}
\end{equation}
This equation can be used as a starting point for RWA approximation.

Consider the case when the only one condition $\delta_{12}^{+}=\epsilon_{1}+\epsilon_{2}+K_{12}^{+}\omega\approx 0$ is satisfied. Nonzero matrix elements of the Hamiltonian $\hat{H}_{2,I}\left(t\right)$ have the following form:
\begin{align}
&\left(\hat{H}_{2,I}\left(t\right)\right)_{11}=
-\frac{1}{4}\Im\left(\Delta_1^2\xi_1^{+}\Xi_1^{+*}+\Delta_2^2\xi_2^{+}\Xi_2^{+*}\right),\nonumber\\
&\left(\hat{H}_{2,I}\left(t\right)\right)_{22}=
-\frac{1}{4}\Im\left(\Delta_1^2\xi_1^{-}\Xi_1^{-*}-\Delta_2^2\xi_2^{+}\Xi_2^{+*}\right),\nonumber\\
&\left(\hat{H}_{2,I}\left(t\right)\right)_{33}=
\frac{1}{4}\Im\left(\Delta_1^2\xi_1^{+}\Xi_1^{+*}-\Delta_2^2\xi_2^{-}\Xi_2^{-*}\right),\nonumber\\
&\left(\hat{H}_{2,I}\left(t\right)\right)_{44}=
\frac{1}{4}\Im\left(\Delta_1^2\xi_1^{-}\Xi_1^{-*}+\Delta_2^2\xi_2^{-}\Xi_2^{-*}\right),\nonumber\\
&\left(\hat{H}_{2,I}\left(t\right)\right)_{14}=
\frac{i}{8}\Delta_1\Delta_2\left(\xi_1^{-}\Xi_2^{+}-\xi_1^{+}\Xi_2^{-}\right.\label{eq:hamiltonian_interaction_representation_2_elements}\\
&\left.\hspace{4cm}+\xi_2^{-}\Xi_1^{+}-\xi_2^{+}\Xi_1^{-}\right)^{*},\nonumber\\
&\left(\hat{H}_{2,I}\left(t\right)\right)_{23}=
\frac{i}{8}\Delta_1\Delta_2\left(\xi_1^{+*}\Xi_2^{+}-\xi_1^{-*}\Xi_2^{-}\right.\nonumber\\
&\left.\hspace{4cm}+\xi_2^{-}\Xi_1^{-*}-\xi_2^{+}\Xi_1^{+*}\right),\nonumber\\
&\left(\hat{H}_{2,I}\left(t\right)\right)_{41}=\left(\hat{H}_{2,I}\left(t\right)\right)_{14}^{*},\nonumber\\
&\left(\hat{H}_{2,I}\left(t\right)\right)_{32}=\left(\hat{H}_{2,I}\left(t\right)\right)_{23}^{*},\nonumber
\end{align}
where the functions $\Xi_q^{\pm}\left(t\right)=\int_{0}^{t}\text{d}\tau\xi_q^{\pm}\left(\tau\right)$ are introduced.
At this point the rotating wave approximation (RWA) is made. Using the generating function of the Bessel functions of the first kind, $e^{iz\sin q}=\sum_k J_k\left(z\right)e^{ikq}$, we can isolate slow oscillations of new Hamiltonian $\hat{H}_{2,I}^{\text{RWA}}\left(t\right)$ obtained from Eq.~\eqref{eq:hamiltonian_interaction_representation_2_elements} by applying RWA looks like
\begin{equation}
\begin{gathered}
\hat{H}_{2,I}^{\text{RWA}}\left(t\right)=\begin{pmatrix}
h_{11} & 0 & 0 & h_{14}e^{-i\Theta\left(t\right)} \\
0 & h_{22} & 0 & 0 \\
0 & 0 & h_{33} & 0 \\
h_{14}e^{i\Theta\left(t\right)} & 0 & 0 & h_{44}
\end{pmatrix}\!,
\end{gathered}
\label{eq:hamiltonian_interaction_2_RWA}
\end{equation}
where
\begin{widetext}
\begin{gather}
h_{11}=-\frac{1}{4}\sum_{k=-\infty}^{+\infty}J_k^2\left(\frac{A}{\omega}\right)\left(\frac{\Delta_1^2}{\epsilon_{1}+g+k\omega}+\frac{\Delta_2^2}{\epsilon_{2}+g+k\omega}\right),\nonumber \quad\\
h_{22}=-\frac{1}{4}\sum_{k=-\infty}^{+\infty}J_k^2\left(\frac{A}{\omega}\right)\left(\frac{\Delta_1^2}{\epsilon_{1}-g+k\omega}-\frac{\Delta_2^2}{\epsilon_{2}+g+k\omega}\right),\nonumber\\
h_{33}=\frac{1}{4}\sum_{k=-\infty}^{+\infty}J_k^2\left(\frac{A}{\omega}\right)\left(\frac{\Delta_1^2}{\epsilon_{1}+g+k\omega}-\frac{\Delta_2^2}{\epsilon_{2}-g+k\omega}\right), \quad\\
h_{44}=\frac{1}{4}\sum_{k=-\infty}^{+\infty}J_k^2\left(\frac{A}{\omega}\right)\left(\frac{\Delta_1^2}{\epsilon_{1}-g+k\omega}+\frac{\Delta_2^2}{\epsilon_{2}-g+k\omega}\right),\nonumber\\
h_{14}=\Delta_1\Delta_2\frac{g}{4}\sum_{k=-\infty}^{+\infty}\!J_k\!\left(\frac{A}{\omega}\right)\!J_{K_{12}^{+}-k}\!\left(\frac{A}{\omega}\right)\!\left(\frac{1}{\left(\epsilon_{1}+k\omega\right)^2-g^2}+\frac{1}{\left(\epsilon_{2}+k\omega\right)^2-g^2}\right),\nonumber \quad
\Theta\left(t\right)=\delta_{12}^{+}t+2\frac{A}{\omega}\sin\varphi_0-K_{12}^{+}\varphi_0.
\label{eq:hamiltonian_interaction_representation_rwa_4_elements}
\end{gather}
At $g\ne0$ we also have
\begin{equation}
\begin{aligned}
\hat{\Gamma}_I^{\text{RWA}}\!&=\!\sum\limits_{q=1}^2\!{\left(\!\Gamma_{\varphi_q}\hat{D}\!\left[\sigma_z^{\left(q\right)}\right]\!+\!\frac{\Gamma_q}{2}\hat{D}\!\left[\sigma_{+}^{\left(q\right)}\right]\!+\!\frac{\Gamma^\prime_q}{2}\hat{D}\!\left[\sigma_{-}^{\left(q\right)}\right]\!\right)} \\
&+\frac{\Gamma_1}{2}\hat{D}\!\left[\sigma_{+}^{\left(1\right)}\sigma_z^{\left(2\right)}\right]\!+\!\frac{\Gamma^\prime_1}{2}\hat{D}\!\left[\sigma_{-}^{\left(1\right)}\sigma_z^{\left(2\right)}\right]+\frac{\Gamma_2}{2}\hat{D}\!\left[\sigma_z^{\left(1\right)}\sigma_{+}^{\left(2\right)}\right]\!+\!\frac{\Gamma^\prime_2}{2}\hat{D}\!\left[\sigma_z^{\left(1\right)}\sigma_{-}^{\left(2\right)}\right]\!.
\end{aligned}
\label{eq:operator_dissipative_interaction_RWA}
\end{equation}
\end{widetext}

For the further analysis it is convenient to get rid of the time dependence in the equation obtained for $\hat{\rho}_{II}$ in the RWA approximation.
For this purpose, let us perform the canonical transformation $\hat{\rho}_{II}=\hat{U}_{\Theta}\hat{\bar{\rho}}_{II}\hat{U}_{\Theta}^{\dag}$ with
\begin{equation}
\hat{U}_\Theta\left(t\right)=\mathrm{diag}\left[e^{-i\Theta\left(t\right)/2},1,1,e^{i\Theta\left(t\right)/2}\right].
\end{equation}
Hence after the made substitution in RWA approximation we have the stationary master equation
\begin{equation}
\frac{\partial\hat{\bar{\rho}}_{II}}{\partial t}=-i\left[\hat{\bar{H}}_{2,I}^{\text{RWA}},{\hat{\bar{\rho} }_{II}}\right]+\hat{\Gamma}_I^{\text{RWA}}{\hat{\bar{\rho}}_{II}},
\label{eq:master_equation_interaction_representation_2_RWA_stationary}
\end{equation}
with Hamiltonian
\begin{equation}
\begin{gathered}
\hat{\bar{H}}_{2,I}^{\text{RWA}}=\begin{pmatrix}
h_{11}-\delta_{12}^{+}/2 & 0 & 0 & h_{14} \\
0 & h_{22} & 0 & 0 \\
0 & 0 & h_{33} & 0 \\
h_{14} & 0 & 0 & h_{44}+\delta_{12}^{+}/2
\end{pmatrix}\!,
\end{gathered}
\label{eq:hamiltonian_interaction_2_RWA_stationary}
\end{equation}
and dissipator defined by Eq.~\eqref{eq:operator_dissipative_interaction_RWA}.

Our analysis of Eq.~\eqref{eq:master_equation_interaction_representation_2_RWA_stationary} shows that there is a unique steady-state:
\begin{equation}
\hat{\bar{\rho}}_{II}=\begin{pmatrix}
r_{11} & 0 & 0 & r_{14} \\
0 & r_{22} & 0 & 0 \\
0 & 0 & r_{33} & 0 \\
r_{41} & 0 & 0 & r_{44}
\end{pmatrix}\!, \\
\label{eq:master_equation_interaction_representation_2_RWA_stationary_steady_state} 
\end{equation}
\begin{align}
r_{11}&=\frac{1}{Z}\!\left[2\frac{\left(\Gamma'_1+\Gamma_2\right)\left(\Gamma_1+\Gamma'_2\right)}{\Gamma_1+\Gamma'_1+\Gamma_2+\Gamma'_2}h_{14}^2\Re{\left(h_g\right)}+\Gamma_1\Gamma_2\left|h_g\right|^2\right]\!, \notag \\
r_{22}&=\frac{1}{Z}\!\left[2\frac{\left(\Gamma_1+\Gamma'_2\right)^2}{\Gamma_1+\Gamma'_1+\Gamma_2+\Gamma'_2}h_{14}^2\Re{\left(h_g\right)}+\Gamma_1\Gamma'_2\left|h_g\right|^2\right]\!, \notag \\
r_{33}&=\frac{1}{Z}\!\left[2\frac{\left(\Gamma'_1+\Gamma_2\right)^2}{\Gamma_1+\Gamma'_1+\Gamma_2+\Gamma'_2}h_{14}^2\Re{\left(h_g\right)}+\Gamma'_1\Gamma_2\left|h_g\right|^2\right]\!, \notag \\
r_{44}&=\frac{1}{Z}\!\left[2\frac{\left(\Gamma'_1+\Gamma_2\right)\left(\Gamma_1+\Gamma'_2\right)}{\Gamma_1+\Gamma'_1+\Gamma_2+\Gamma'_2}h_{14}^2\Re{\left(h_g\right)}+\Gamma'_1\Gamma'_2\left|h_g\right|^2\right]\!, \notag \\
r_{14}&=r_{41}^{*}=i h_{14}\frac{\Gamma_1\Gamma_2-\Gamma'_1\Gamma'_2}{Z}{h_g}, \label{eq:rij}
\end{align}
where constants $h_g$ and $Z$ are equal respectively
\begin{align}
h_g\!&=\!2\Gamma_{\varphi_1}\!+\!\frac{\Gamma_1\!+\!\Gamma'_1}{2}\!+\!2\Gamma_{\varphi_2}\!+\!\frac{\Gamma_2\!+\!\Gamma'_2}{2}\!-\!i\left(h_{11}\!-\!h_{44}\!-\!\delta_{12}^{+}\right), \notag \\
Z&=\left(\Gamma_1+\Gamma'_1\right)\left(\Gamma_2+\Gamma'_2\right){\left|h_g\right|^2} \\
&\hspace{20mm}+2h_{14}^2\left(\Gamma_1+\Gamma'_1+\Gamma_2+\Gamma'_2\right)\Re{\left(h_g\right)}. \notag
\end{align}
It is known that an unique steady-state of the LME is attractive, i.e., all other solutions converge to it~\cite{Schirmer2010}. Going back to the original representation, we find that in the main order, the steady-state of the system ($t\to\infty$) is defined as:
\begin{equation}
\hat{\rho}=\begin{pmatrix}
r_{11} & 0 & 0 & r_{14}e^{-i\phi} \\
0 & r_{22} & 0 & 0 \\
0 & 0 & r_{33} & 0 \\
r_{41}e^{i\phi} & 0 & 0 & r_{44}
\end{pmatrix}\!,
\label{eq:master_equation_stationary_steady_state}
\end{equation}
where $\phi=K_{12}^{+}\omega t-2\frac{A}{\omega}\sin\left(\omega t-\varphi_0\right)-K_{12}^{+}\varphi_0$.

\section{Concurrence in the resonant case}
The solution Eq.~\eqref{eq:master_equation_stationary_steady_state} allows the calculation of concurrence when the resonance condition $\delta_{12}^{+}=\epsilon_{1}+\epsilon_{2}+K_{12}^{+}\omega\approx 0$ is met. It is easy to show that the eigenvalues of the corresponding matrix $R$ are equal to
\begin{equation}
\begin{aligned}
\lambda_1&=\sqrt{r_{11}r_{44}}+\left|r_{14}\right|, \\
\lambda_2&=\left|\sqrt{r_{11}r_{44}}-\left|r_{14}\right|\right|, \\
\lambda_3&=\lambda_4=\sqrt{r_{22}r_{33}},
\end{aligned}
\label{eq:resonance_R_eigenvalues}
\end{equation}
where the values $r_{ij}$ are defined by Eq.~\eqref{eq:rij}. A more detailed analysis of the eigenvalues Eq.~\eqref{eq:resonance_R_eigenvalues} shows that $\lambda_1\ge\lambda_2,\lambda_3,\lambda_4$. Indeed, by virtue of the fact that $\Re{\left(h_g\right)}\ge\left(\Gamma_1\!+\!\Gamma'_1\!+\!\Gamma_2\!+\!\Gamma'_2\right)/2$, the following chain of inequalities holds:
\begin{equation}
\begin{aligned}
r_{11}r_{44}\ge\frac{1}{Z^2}&\left[\left(\Gamma'_1+\Gamma_2\right)\left(\Gamma_1+\Gamma'_2\right)h_{14}^2+\Gamma_1\Gamma_2\left|h_g\right|^2\right] \\
\times&\left[\left(\Gamma'_1+\Gamma_2\right)\left(\Gamma_1+\Gamma'_2\right)h_{14}^2+\Gamma'_1\Gamma'_2\left|h_g\right|^2\right] \\
=\left|r_{14}\right|^2+\frac{1}{Z^2}&\left[\left(\Gamma'_1+\Gamma_2\right)^2h_{14}^2+\Gamma'_1\Gamma_2\left|h_g\right|^2\right] \\
\times&\left[\left(\Gamma_1+\Gamma'_2\right)^2h_{14}^2+\Gamma_1\Gamma'_2\left|h_g\right|^2\right],
\end{aligned}
\end{equation}
where it follows that $r_{11}r_{44}\ge\left|r_{14}\right|^2$ and, therefore, $\lambda_2=\sqrt{r_{11}r_{44}}-\left|r_{14}\right|$,
which proves the inequality $\lambda_1\ge\lambda_2$. It can also be checked by the simple comparison that $r_{11}r_{44}\ge{r_{22}}r_{33}$, whence $\lambda_1\ge\lambda_3,\lambda_4$. Thus,
\begin{equation}
\overline{C}=\max\left(0,2\left(\left|r_{14}\right|-\sqrt{r_{22}r_{33}}\right)\right).
\label{eq:C_result_resonance}
\end{equation}

The expression~\eqref{eq:C_result_resonance} allows us to estimate the width of the concurrence value $\overline{C}$ dips.
Thus, when the relation $\epsilon_2=s\epsilon_1$ is fulfilled, the width of the dip on the parameter $\epsilon_1$ is equal to
\begin{equation}
\begin{aligned}
\Delta\epsilon_1=\frac{2}{1+s}\left(\frac{\Gamma_1+\Gamma_2}{2}+2 (\Gamma_{\varphi_1}+\Gamma_{\varphi_2})\right)\\
\times\sqrt{\left(\frac{2h_{14}}{\Gamma_1+\Gamma_2}\right)^2-1}. \label{eq:35}
\end{aligned}
\end{equation}
Hence we can conclude that in the presence of dephasing ($\Gamma_{\varphi_1}+\Gamma_{\varphi_2}>0$), nonzero entanglement is possible only in the presence of dissipation ($\Gamma_1+\Gamma_2>0$). In the absence of dephasing ($\Gamma_{\varphi_1}+\Gamma_{\varphi_2}=0$) entanglement is possible in both dissipative and non-dissipative cases.

\section{RESULTS AND DISCUSSION}
\begin{figure}[b]
  \includegraphics[width=1.\columnwidth]{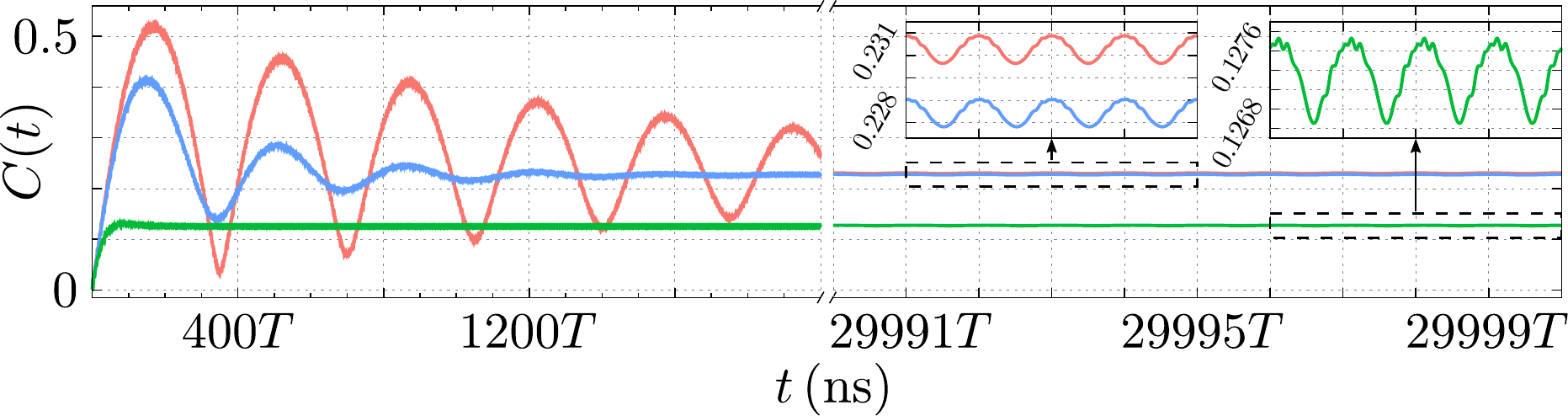}
  \caption{Concurrence dynamics $C\left(t\right)$ on long time interval. The results of direct numerical calculations are presented for $\Gamma_1=\Gamma_2=0.0001$~GHz (red curve), $\Gamma_1=\Gamma_2=0.0005$~GHz (blue curve) and $\Gamma_1=\Gamma_2=0.005$~GHz (green curve). The insets show the dependence $C\left(t\right)$ on an enlarged scale. Other parameters: $\Delta_1=0.1$~GHz, $\Delta_2=0.15$~GHz, $\epsilon_1=3.331$~GHz, $\epsilon_2=6.662$~GHz, $g=0.15$~GHz, $A=5$~GHz, $\omega=1$~GHz ($T=2\pi/\omega$), $\Gamma_{\varphi_1}=\Gamma_{\varphi_2}=0$~GHz, $T_1=T_2=30$~mK.} \label{fig1}
\end{figure}

First, we numerically analyzed the temporal dynamics of the concurrence, $C(t)$, of two interacting qubits under different dissipation parameters $\Gamma_{1,2}$, see Fig.~\ref{fig1}. Under the action of classical control fields, $\epsilon_q(t)$, the system of coupled qubits in independent reservoirs is characterized by a finite degree of entanglement. It is seen that the formation of steady-state essentially depends on the dissipation rate in the system (the higher the dissipation rate, the faster the steady state is formed). The long-lived stable entangled state maintains a periodic dependence with the oscillation frequency equal to the frequency of external influence $\omega$, as shown in the insets on Fig.~\ref{fig1}, and does not depend on the initial state. Such behavior is fully consistent with the analytical expression~\eqref{eq:C_result_no_resonance}. 

\begin{figure}[t]
  \includegraphics[width=1\columnwidth]{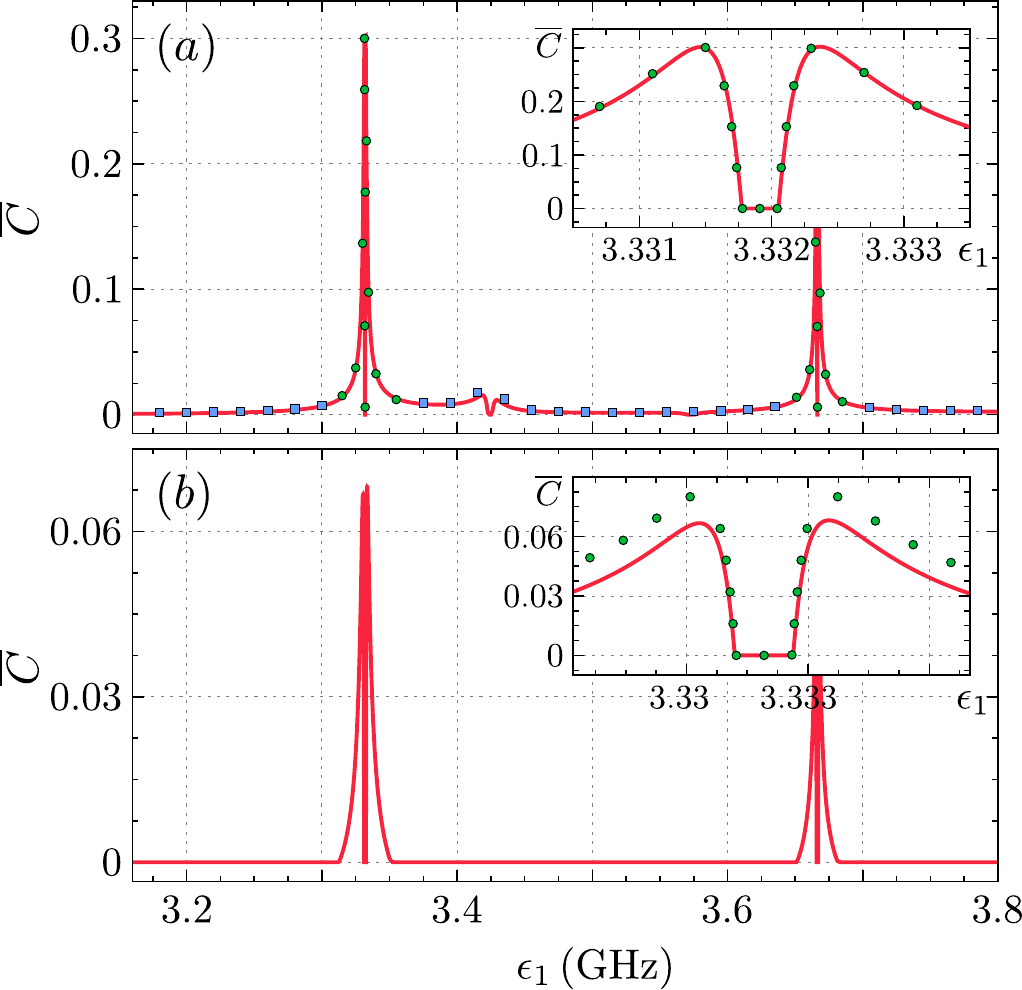}
  \caption{Average concurrence $\overline{C}$ as functions of the control
parameter $\epsilon_1$ for different noise rates: (a) $\Gamma_{\varphi_1}=\Gamma_{\varphi_2}=0$~GHz, $\Gamma_1=\Gamma_2=10^{-4}$~GHz (b) $\Gamma_{\varphi_1}=\Gamma_{\varphi_2}=10^{-4}$~GHz, $\Gamma_1=\Gamma_2=10^{-4}$~GHz. The results of direct numerical calculations (solid curves) compared with analytical results: square markers are derived from non-resonant expressions~\eqref{eq:C_result_no_resonance_average}; round ones are derived from resonant expression~\eqref{eq:C_result_resonance}. The following parameters were used: $\Delta_1=0.1$~GHz, $\Delta_2=0.15$~GHz, $g=0.15$~GHz, $A=5$~GHz, $\omega=1$~GHz, $T_1=T_2=30$~mK, $\epsilon_2=2\epsilon_1$.} \label{fig2}
\end{figure}

\begin{figure}[t]
  \includegraphics[width=1\columnwidth]{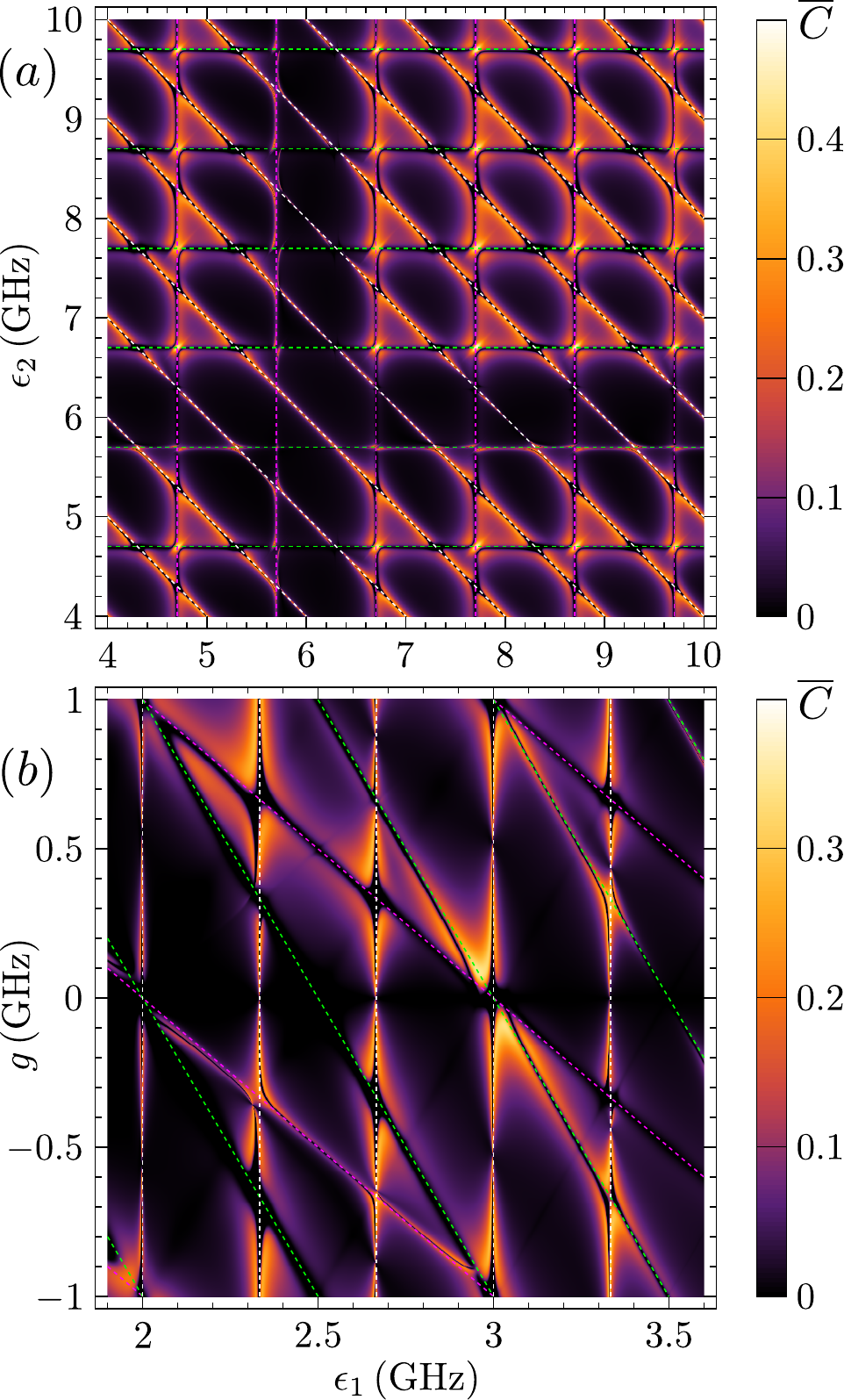}
  \caption{Average concurrence $\overline{C}$ as function of the control
parameter $\epsilon_1$, $\epsilon_2$ and $g$. The following parameters were used: (a) $g=0.3$~GHz, $A=10$~GHz; (b) $A=5$~GHz, $\epsilon_2=2\epsilon_1$. Dotted lines correspond to the resonance conditions~\eqref{eq:resonance_condition_a} (purple and green lines) and~\eqref{eq:resonance_condition_b} (white lines). Other parameters: $\Delta_1=0.2$~GHz, $\Delta_2=0.3$~GHz, $\omega=1$~GHz, $\Gamma_{\varphi_1}=\Gamma_{\varphi_2}=0$~GHz, $\Gamma_1=\Gamma_2=10^{-3}$~GHz, $T_1=T_2=30$~mK.} \label{fig3}
\end{figure}

Naturally, the external field and dissipation parameters have a significant influence on the values of concurrence in the long-lived steady-state into which the information is encoded. Therefore, to further entanglement and state control investigation, we will study the behavior of the concurrence, $\overline{C}$, averaged over the pulse length $\tau$ and initial phase $\varphi_0$. Figure~\ref{fig2} demonstrates the comparison of direct numerical simulation results (solid curves in Fig.~\ref{fig2}) as a function of the static bias,  component of external field, $\epsilon_1$, acting on the first qubit, with the analytical expressions found outside the Eq.~{\eqref{eq:C_result_no_resonance_average}} resonances (square markers in Fig.~\ref{fig2}) and near the Eq.~\eqref{eq:C_result_resonance} resonances (round markers in Fig.~\ref{fig2}). Note that a good agreement of the obtained results is observed, including the formation of entanglement breaking regions near the fulfillment of the conditions~\eqref{eq:resonance_condition_a} and~\eqref{eq:resonance_condition_b}. The effect of abrupt suppression of concurrence (falling $\overline{C}$ value to zero) is known as sudden death of entanglement~\cite{Yu2009,Obada2012} and is associated with uncorrelated weak reservoir noise. Such a phenomenon has previously been experimentally demonstrated in atomic and photonic systems~\cite{Almeida2007,Yu2020}, and has important practical implications for manipulating entangled states on demand~\cite{Verstraete2009}. From the resonance perturbation theory developed in Sec.~V, we estimated the magnitudes of the dips in the $\overline{C}(\epsilon_1)$ dependence. From the expression~\eqref{eq:35}, it can be seen how the magnitudes of the dips depend on the noise in the system.

\begin{figure}[b]
  \includegraphics[width=1\columnwidth]{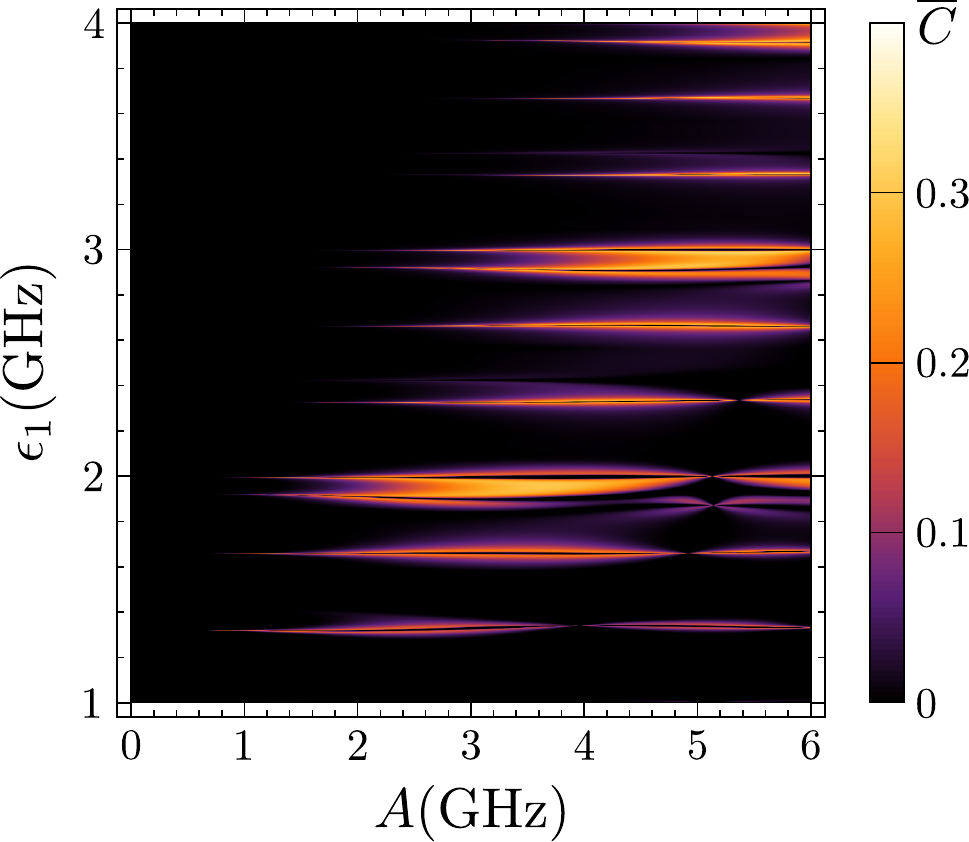}
  \caption{Average concurrence $\overline{C}$ as function of the control
parameter $A$ and $\epsilon_1$. Parameters: $\Delta_1=0.2$~GHz, $\Delta_2=0.3$~GHz, $g=0.15$~GHz, $\omega=1$~GHz, $\Gamma_{\varphi_1}=\Gamma_{\varphi_2}=0$~GHz, $\Gamma_1=\Gamma_2=10^{-3}$~GHz, $T_1=T_2=30$~mK, $\epsilon_2=2\epsilon_1$.}  \label{fig4}
\end{figure}

To understand in detail how the entanglement degree in the system depends on the parameters of the external control fields and the coupling strength between the qubits, color maps for the dissipatively stable long-lived steady-state were calculated, as shown in Fig.~\ref{fig3} and Fig.~\ref{fig4}. It can be noticed that structurally these dependences coincide with the regions of formation of multiphoton transitions at LZSM interference found in our early works~\cite{Satanin2012, Munyaev2021}. For clarity, the dashed lines in Fig.~\ref{fig3} show the conditions of resonances~\eqref{eq:resonance_condition_a} (purple and green lines) and~\eqref{eq:resonance_condition_b} (white lines), under which the effect of entanglement sudden death occurs. Bright regions near the resonances correspond to the maximum degree of entanglement in the system. Thus, we find that quantum entanglement and nonlocality in the system can be manipulated and controlled by performing tuning of the external fields amplitudes, $\epsilon_{1,2}$, $A$.


\section{CONCLUSION}

In this paper, we study the formation of stationary entanglement at LZSM interference in a system of two inductively coupled flux qubits, each coupled to its own local bosonic reservoir, using the basic Floquet--Markov equation. Analytical expressions for averaged stationary concurrence are obtained, which in combination with numerical simulations allow us to study in detail the behavior of entanglement depending on the parameters of external influence and dissipation. It is demonstrated that the qubits coupling to the environment leads to the formation of dissipatively stable long-lived steady-state. The process of generation and destruction of these states, including the effect of entanglement sudden death, by tuning the amplitudes of bias and external driving fields is proposed~\cite{Yu2009, Obada2012}. This technique can be used in the design of quantum circuits to collect and propagate stationary entanglement where it is needed in a quantum processor, such as when transferring information between nodes in a quantum network~\cite{Kraus2004}. In addition, dissipative state preparation can be used to create complex entangled states on demand, a process that is seen as a form of adiabatic quantum computing~\cite{Verstraete2009}.

\section{ACKNOWLEDGMENTS}
The work was supported by UNN within the framework of the strategic academic leadership program ``Priority 2030'' of Ministry of Science and Higher Education of the Russian Federation.

\bibliographystyle{apsrev4-1}
\bibliography{bibliography}

\end{document}